\documentclass[preprint]{aastex}
\usepackage{emulateapj5}
\begin{document}
\title{Extraordinary Late-Time Infrared Emission of Type~IIn 
Supernovae \footnotemark}

\shorttitle{Dust Emission in SNe~IIn}
\shortauthors{Gerardy et al.}
  
\author{Christopher L. Gerardy,\altaffilmark{2} Robert A. Fesen,\altaffilmark{2}
Ken'ichi Nomoto,\altaffilmark{3} Peter M. Garnavich,\altaffilmark{4} 
Saurabh Jha,\altaffilmark{5} Peter M. Challis,\altaffilmark{5} 
Robert P. Kirshner,\altaffilmark{5} Peter H\"{o}flich,\altaffilmark{6} and 
J. Craig Wheeler\altaffilmark{6}} 
\footnotetext{Based in part on data collected at Subaru Telescope, which is
operated by the National Astronomy Observatory of Japan}
\altaffiltext{2}{Department of Physics and Astronomy, 6127 Wilder Laboratory,
Dartmouth College, Hanover, NH 03755-3528}
\altaffiltext{3}{Department of Astronomy and Research Center for the Early 
Universe, University of Tokyo, Bunkyo-ku, Tokyo 113-0033, Japan}
\altaffiltext{4}{Department of Physics, University of Notre Dame, 225 Nieuwland 
Science Hall, Notre Dame, IN 46556}
\altaffiltext{5}{Harvard-Smithsonian Center for Astrophysics, 60 Garden Street,
Cambridge, MA 02138}
\altaffiltext{6}{Department of Astronomy, University of Texas at Austin,
Austin, TX 78712}

\begin{abstract}
Near-Infrared (NIR) observations are presented for five Type~IIn supernovae 
(SN~1995N, SN~1997ab, SN~1998S, SN~1999Z, and SN~1999el) that exhibit strong 
infrared excesses at late times (t~$\ga 100$~d).  \textit{H}- and 
\textit{K}-band emission from these objects is dominated by a continuum that 
rises toward longer wavelengths.  The data are interpreted as thermal emission 
from dust, probably situated in a pre-existing circumstellar nebula.  The IR 
luminosities implied by single temperature blackbody fits are quite large, 
$\ga 10^{41 - 42}$ erg~s$^{-1}$, and the emission evolves slowly, lasting for 
years after maximum light.  For SN~1995N, the integrated energy release via IR 
dust emission was $\approx 0.5$ -- $1 \times 10^{50}$ erg.  A number of dust 
heating scenarios are considered, the most likely being an infrared echo powered
by X-ray and UV emissions from the shock interaction with a dense circumstellar 
medium.
\end{abstract}

\keywords{supernovae: individual (SN~1995N, SN~1997ab, SN~1998S, SN~1999Z, 
SN~1999el) --- dust --- circumstellar matter --- infrared: stars}

\section{Introduction}
\subsection{Background}
On 3 January 1980, \citet{merrill80} detected infrared emission from SN~1979C 
that was stronger toward longer wavelengths.  Since then similar ``infrared
excesses'' have been observed in a number of other supernovae including 
SN~1980K \citep{telesco81}, SN~1982E \citep{graham83,graham86}, SN~1982L, 
SN~1982R \citep{graham86}, SN~1985L \citep{elias86}, SN~1993J \citep{lewis94},
SN~1994Y \citep{garnavich96},  SN~1998S \citep{gerardy00,fassia00} and 
SN~1997ab \citep{gerardy00}.  These infrared excesses have generally been 
interpreted as thermal emission from dust, either forming in the ejecta
\citep{merrill80,telesco81,dwek83a,elias86,kozasa89a,kozasa89b,kozasa91,lucy91}
or lying in a pre-existing circumstellar medium and heated by the supernova 
flash as an ``infrared echo''
\citep{bode80,dwek83b,graham83,graham86,lewis94,garnavich96}. 

In the first scenario, dust forms in the cooling ejecta after it reaches a 
critical temperature.  Late time thermal emission from such dust has been 
invoked many times to explain NIR excesses observed in nova outbursts 
\citep{gerhz88}, where the critical formation temperature is typically 
1000--1200 K.  While such a picture is perhaps the simplest explanation for the 
dust emission (e.~g., SN~1987A: 
\citealt{kozasa89a,kozasa89b,kozasa91,danziger91,lucy91}), it can often be ruled
out by energy considerations.  Since the dust grain cooling time is short and 
dust is localized in the ejecta where light travel times are small, the infrared
luminosity must be limited by the energy available in the supernova at the time 
of observation. These sources include the thermal energy stored in the dusty 
gas, UV/X-ray radiation from the supernova, and gamma radiation from radioactive
heavy elements in the ejecta.  At late times ($t \geq 200$~d) the luminosity of 
observed IR excesses becomes increasingly difficult to explain within the dust 
formation scenario \citep{dwek83b,gerardy00}.

In light of this problem, a number of authors have suggested that the observed
IR excesses are due to infrared dust echos.  In this scenario, the emitting dust
is not localized to the ejecta, but is distributed in an extended circumstellar 
nebula.  As the supernova flash propagates outward into the nebula, the dust is 
heated and then radiates thermally in the infrared.  Thus the late-time IR 
luminosity is powered by the maximum-light flash which, even after geometric 
dilution over a large sphere, is potentially a larger energy source than that 
available to dust in the SN ejecta.  

Such energy considerations tended to make IR echos a more attractive explanation
for the observed infrared excesses.  For example, \citet{dwek83b} found that 
the IR emission from the Type~IIL events SN~1979C and SN~1980K was too bright, 
and that there was not enough energy in these objects at late times for a dust 
formation scenario.  They were, however, able to fit the observed NIR excesses 
with an echo model.  Similarly, \citet{graham86} were able to use a IR echo 
model to reproduce the observed IR light curve of the untyped object SN~1982E. 
 
On the other hand, \citet{elias86} observed IR emission in SN~1985L that became 
brighter before fading, in contradiction to IR echo models that predict a 
plateau followed by fading and only rise at very early times when the supernova 
is bright.  \citet{elias86} argued that this was therefore a strong case for 
dust formation rather than an infrared echo.  However, \citet{emmering88} 
published models of echos in aspherical dust clouds which showed that IR echos 
from highly flattened dust distributions could exhibit more complicated rising 
and falling IR light curves.  

Much stronger evidence for dust formation was
seen in SN~1987A where increased infrared emission was accompanied by a 
corresponding decrease in optical emission, while at the same time emission line
profiles were observed to shift toward the blue 
\citep[][and references therein]{mccray93}. Both effects argue in favor of dust 
formed with the supernova ejecta.  Similarly, in SN~1998S, the evolution of the hydrogen 
and helium line
profiles also suggested dust formed in the ejecta as
the red-shifted side of the profile steadily faded while the blue-shifted side
remained nearly constant \citep{gerardy00}.  In this case, however, the thermal 
emission in the NIR was too bright to have come solely from the ejecta 
\citep{fassia00} suggesting that there may be two different dust components
in SN~1998S: newly formed dust in the ejecta and brightly emitting dust in the
dense CSM. 

\subsection{Recent Results On Type~IIn Supernovae}
The NIR excesses observed in SN~1998S and SN~1997ab have added new dimensions 
to the problem \citep{gerardy00,fassia00}.   
Both objects are classified as Type~IIn 
events, a sub-class of Type~II supernovae that exhibit multi-component 
``Seyfert-like'' hydrogen and helium emission lines \citep{schlegel90}.  This 
class is often associated with very slowly decaying light curves and strong 
X-ray emission \citep[and references therein]{filippenko97}.  They are thought 
to be interacting with a dense and perhaps clumpy circumstellar wind resulting
from significant mass-loss prior to core-collapse \citep{chugai94}.

The IR emission in SN~1997ab and SN~1998S was quite luminous 
($10^{40-41}$ erg~s$^{-1}$) at late epochs.  For comparison, the excesses 
seen in SN~1979C and SN~1980K were first observed around day 200 and decayed 
exponentially with e-folding times $\sim 100$ days. Yet the strong NIR excess 
in SN~1997ab was observed more than three years after the explosion. 

Large IR luminosities at late epochs strained both the dust formation and IR 
echo models, especially for SN~1997ab.  Furthermore, the NIR spectra of 
SN~1997ab and SN~1998S were nearly identical even though SN~1997ab was 
approximately three times older.  This seemed to imply either that these events 
were coincidentally observed when they would show the same spectrum, or that 
the evolutionary time scale of these objects was quite long. 
These issues, plus the Type~IIn classification of SN~1997ab and SN~1998S, led 
\citet{gerardy00} to propose that the IR emission was 
powered by the circumstellar interaction either directly through shock heating
or via UV/X-ray emission from the shocked gas.  They were unable to distinguish 
between circumstellar dust and newly formed ejecta dust as the 
emission source, 
but \citet{fassia00} used \textit{K} and \textit{L$^\prime$} photometry from 
an earlier epoch to rule out ejecta dust as the principle NIR emission source 
in SN~1998S.

In this paper, we present further near-infrared observations following the 
evolution of SN~1998S and SN~1997ab. We also present NIR observations of 
three other Type~IIn supernovae, SN~1995N, SN~1999Z, and SN~1999el, which 
also exhibit large late-time IR excesses.   In \S 2, we discuss the
observations and data reduction procedures.  Near-IR spectral 
energy distributions are presented for all objects in \S 3, along with
NIR spectra of SN~1995N at four epochs covering nearly five years of evolution.
We then consider several emission scenarios, including 
interaction shock precursor, shock heated dust, simple IR echo, and interaction 
powered echo models.  We summarize our conclusions in \S 4.

\section{Observations}
\subsection{Supernova Dates and Distances}
SN~1995N was discovered on 5 May 1995 by \citet{pollas95} in MCG-02-38-017.
Based on optical spectra \citet{benetti95} estimated that it was at least 
10 months old and, following \citet{fransson01}, we will adopt 4 July 1994
as the date of zero epoch.  

SN~1998S was discovered in NGC~3877 by Z. Wan 
\citep{li98,qiu98} on 2 March 1998.  Following \citet{gerardy00}, we chose the 
date of maximum light of 18 March 1998 (Garnavich 1998, private communication) 
as the date of zero epoch.  

The other three supernovae were discovered either
near or after maximum light and we adopt the discovery date as zero epoch for
these objects.  SN~1997ab was discovered in an objective prism image taken 
on 11 April 1996 for the Hamburg Quasar Survey, \citep{hagen97b}.  The host
galaxy is the faint dwarf irregular galaxy HS 0948+2018.  

SN~1999Z was
discovered on 8 Feb 1999 by \citet{schwartz99} in UGC~5608, and SN~1999el
was discovered in NGC~6951 by the BAO Supernova Survey on 20 October 1999
\citep{cao99}. 

Approximate distances to the supernovae were calculated from the recession
velocities of the host galaxies, using H$_0 = 72$ km~s$^{-1}$~Mpc$^{-1}$ 
\citep{freedman01}.  For the three nearest objects, SN~1995N, SN1998S, and
SN~1999el, recession velocities corrected for Virgo-centric infall (taken from
the Lyon-Meuden Extragalactic Database\footnote{http://leda.univ-lyon1.fr/}) 
were used.  The resulting distances are 26 Mpc (1857 km~s$^{-1}$), 15.5 Mpc
(1114 km~s$^{-1}$), and 24 Mpc (1705 km~s$^{-1}$) respectively.  For the 
more distant objects SN~1997ab and SN~1999Z, raw observed redshifts of 
3600 km~s$^{-1}$ \citep{hagen97a} and 15100 km~s$^{-1}$ \citep{jha99} 
were used yielding distances of 50 Mpc and 210 Mpc respectively.

\subsection{Near-Infrared Photometry} 
Near-infrared \textit{J}-, \textit{H}-, and \textit{K}-band images of SN~1998S, 
SN~1997ab, SN~1999Z, and SN~1999el were obtained from June 1998 to November 2000
using TIFKAM on the 2.4~m Hiltner telescope at MDM Observatory.  TIFKAM is a 
high-throughput infrared imager and spectrograph with a $512 \times 1024$ 
ALLADIN InSb 
detector.   Dithered sets of on target images were bracketed with images of 
nearby empty fields for sky subtraction.  Most nights were photometric and
photometric solutions were achieved by observing HST/2MASS photometric 
standards \citep{persson98} at a variety of airmasses throughout the night.
The data from non-photometric nights was calibrated by matching the magnitudes 
of field stars to images from photometric nights.  

\textit{J}-, \textit{H}-, and \textit{K}-band images of SN~1995N were taken 
during the period from July 1996 to May 2001.  The majority of observations were
performed on the 1.2~m telescope at F. L. Whipple Observatory (FLWO) and with 
TIFKAM on the 2.4~m Hiltner telescope at MDM Observatory.  Additional 
observations were obtained using the NASA Infrared Telescope Facility (IRTF), 
Apache Point Observatory (APO), and IRCS on the Subaru telescope.  In order to
avoid mixing data fluxed with different photometric systems, all SN~1995N 
observations were fluxed relative to several nearby field stars.  A photometric
solution for the field stars was obtained by observing \citet{persson98} 
standards from MDM.  Thus, the SN~1995N data is on the same photometric system
as the data from the other four supernovae.

\subsection{Near-Infrared Spectroscopy}
Near-infrared spectra of SN~1995N were observed from April 1996 to March 2001.  
\textit{H}- and \textit{K}-band spectra were obtained on April 4 and 5 1996
using FSPEC on the Multiple-Mirror Telescope (MMT), and using the Infrared 
Cryogenic Spectrometer (CRSP) on the 4m Mayall telescope at Kitt Peak National 
Observatory (KPNO) on 16 June 1997 and 12 February 1998.  Each night of the MMT 
data was reduced separately and then combined into a single spectrum.  An 
additional \textit{K}-band spectrum of SN~1995N was observed on 5 March 2001 
using the Infrared Camera and Spectrograph (IRCS; \citealt{kobayashi00}) on the 
Subaru Telescope.  A log of all the near-infrared spectroscopy is presented in 
Table~\ref{nirspec}. 

The spectroscopic observations were taken in many short exposures.  Between
exposures the target was dithered across the slit to sample the chip at multiple
locations and provide first-order background subtraction.  Then 1D spectra were 
extracted using standard IRAF tasks.  Wavelength calibration was achieved
using OH lines in the night sky background for the MMT and KPNO spectra, and 
using arc-lamps for the Subaru data.  

A nearby F5V star from the Bright Star Catalog
\citep{hoffleit82} was used as a telluric standard for the MMT and KPNO data.  
Subaru data was corrected for telluric absorption using nearby AV and GV 
stars selected from the Gemini Spectroscopic Standard Star 
Catalogues\footnote{http://www.us-gemini.noao.edu/sciops/instruments/niri/NIRIIndex.html}.  
The G-dwarf was divided by a solar spectrum 
\citep{livingston91,wallace93}\footnote{NSO Kitt Peak FTS data used here were 
produced by NSF/NOAO} to remove stellar features.  This was used to correct for 
telluric absorption in the A dwarf spectrum.  The stellar features were fit 
from the corrected A-star spectrum and removed from the raw A-star spectrum.  
The raw A-star spectrum with the stellar features removed was then used to 
correct the target data for telluric absorption.  (See 
\citealt{hanson96,hanson98} for a more detailed discussion of this technique.) 
The instrumental response for all the runs was 
removed by matching the observed spectra of the telluric standards to model 
spectra from the stellar atmosphere calculations of \citet{kurucz94} 

The absolute flux levels of the spectra were set by matching the average flux
in each pass band to the NIR photometric data.  For two of the epochs 
(days 1076 and 2435), photometric images were obtained within a few days and 
were used to set the flux.  For the other two epochs, the NIR photometry was 
extrapolated using the two nearest photometric points for each band.  In all 
cases, the pass bands of \citet{persson98} were assumed.  We estimate the flux 
calibration is accurate to about $20\%$.

There are a few issues that we have not accounted for which
could slightly alter the results of our analysis.  For example, the local
reddening of the supernovae is not known for many of these objects and thus no
correction has been applied.  We have also ignored the
potential problems of exact photometric calibration of objects which have 
spectral energy distributions (SEDs) quite different from those of the 
photometric standards; in this case very red continua in
the \textit{H}- and \textit{K}-bands and strong line emission in
\textit{J}-band.  In light of these issues, we have emphasized the qualitative
results from our analysis and treat the quantitative results as 
approximations rather than exact measurements.
 
\section{Analysis}
\begin{figure*}
\epsscale{.70}
\plotone{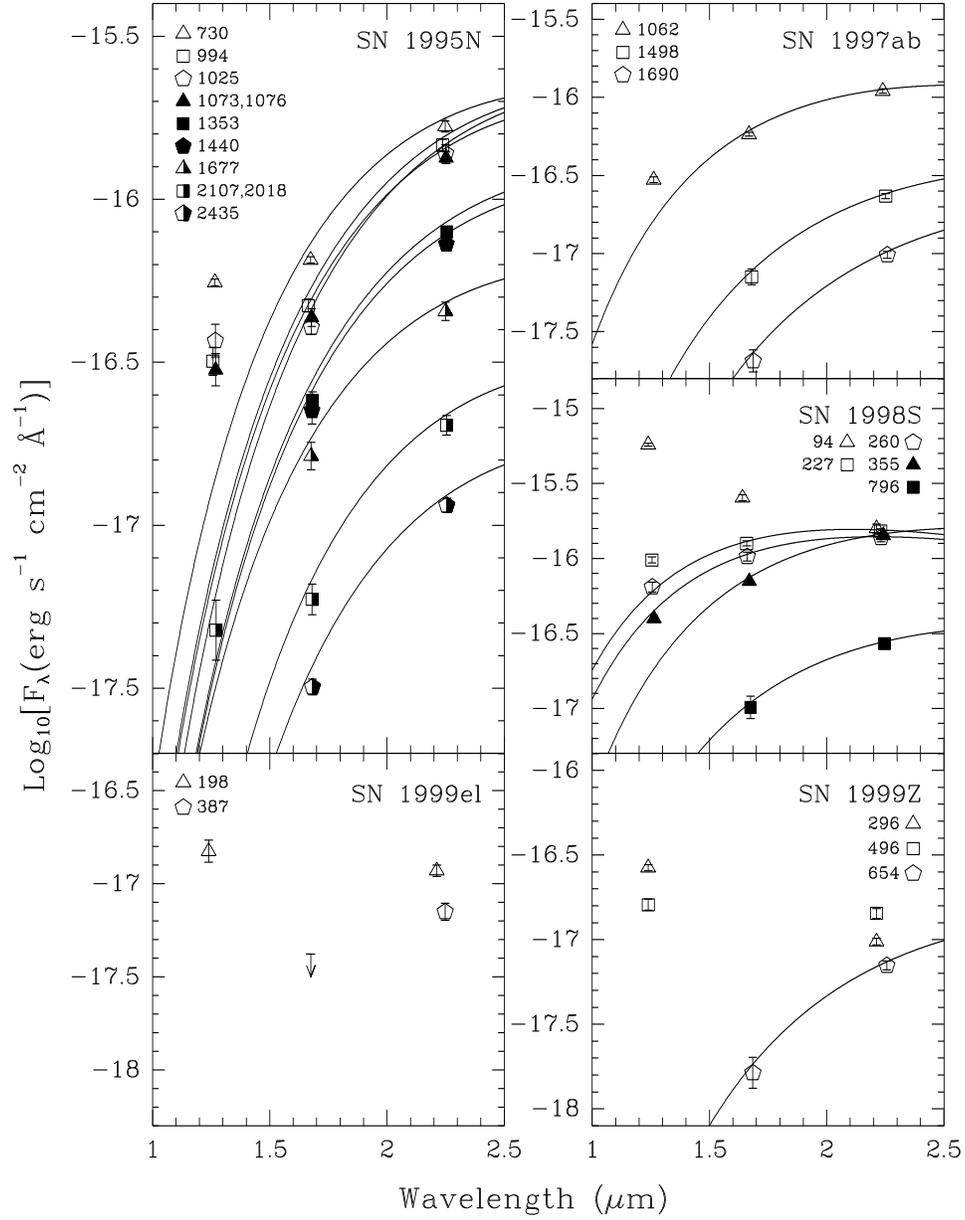}
\caption{Spectral energy distributions for SN~1995N, SN~1997ab, SN~1998S, 
SN~1999Z, and SN~1999el.  The curves are the best fit single temperature
blackbody curves for the observed $H - K$ colors, assuming a $\lambda^{-1}$
absorption/emission efficiency. \label{allsed}}
\end{figure*}

Table~\ref{nirphot} lists our infrared photometric results.  The observed 
magnitudes have been transformed into fluxes using the zero points of 
\citet{beckwith76} and \citet{schultz01}.
Near-infrared spectral energy distributions are plotted for each of 
the supernovae in Figure~\ref{allsed}.  The SEDs have been corrected for 
galactic extinction using the measurements of \citet{schlegel98} and assuming
an $R_V=3.1$ extinction curve.  Not all of the data for SN~1995N is plotted.
Two epochs (days 1375 and 1785) had very poor \textit{H}-band detections and
are omitted in Figure~\ref{allsed}, as are all epochs with only a 
\textit{K}-band detection.

The overall trend is a slow evolution toward redder and fainter 
near-infrared continua.  In three of the objects (SN~1998S, SN~1999Z, and 
SN~1999el) we observe the onset of the NIR excess.  For example,
in SN~1998S, the \textit{K}-band flux remains essentially constant from
day 94 to day 355 while the \textit{J}- and \textit{H}-bands fade as if the
blue photospere fades to reveal a much more slowly evolving red continuum 
underneath.  In contrast, the \textit{K}-band flux increases during the 
blue-to-red transition in SN~1999Z suggesting that the IR continuum might
still have been rising during this period.  SN~1995N and SN~1997ab already
exhibited strong NIR excesses when they were first observed.  During the
observed period, SN~1997ab slowly evolved to redder colors while SN~1995N 
exhibited virtually no color evolution over nearly 1800 days.
 
\subsection{Single-Temperature Fits}
For data showing clear infrared excesses, we fit $H - K$ color temperatures
with dust emission/absorption efficiencies $Q_\lambda = $ constant, 
$Q_\lambda \propto \lambda^{-1}$, and $Q_\lambda \propto \lambda^{-2}$. The results are listed in 
Table~\ref{singletemp}.  We note that single-temperature dust emission
is probably not a good description for the IR excesses, which would generally
be a superposition of many dust temperatures.  (It would be a valid
description if the emitting dust was confined to a thin shell or very far
away from the SN light source so that $r^{-2}$ variations would be small.)
However, since we only have NIR data it is not possible to constrain the
long-wavelength dust emission spectrum.  Thus, our data is really only sensitive
to the hottest components of the dust emission.

The best-fit curves for $Q_\lambda \propto \lambda^{-1}$ are shown in 
Figure~\ref{allsed}.  We note that the \textit{J}-band flux always lies 
significantly above the continuum fit by the $H-K$ color.  IR dust emission is 
only a very small component of the \textit{J}-band flux which is likely 
dominated by the Pa$\beta$ emission line.  As a result, we ignored the 
\textit{J}-band flux in the analysis of the dust emission.  Also, while
we have plotted the $Q_\lambda \propto \lambda^{-1}$ curves, the $Q_\lambda = $
constant and $Q_\lambda \propto \lambda^{-2}$ curves also fit the \textit{H}- 
and \textit{K}-band spectra of SN~1995N.  
Without longer wavelength observations we cannot constrain the dust 
absorption/emission efficiency. For the rest of the analysis we 
will assume $Q_\lambda \propto \lambda^{-1}$ unless otherwise noted. 

The observed $H - K$ temperature evolution of SN~1995N, SN~1997ab, and
SN~1998S are plotted in the upper panels of Figure~\ref{templum}.  SN~1995N
exhibits almost no temperature evolution, cooling only from about 900~K to
about 800~K between
the first and last epochs. In contrast, the other two supernovae show 
significant color evolution.  SN~1997ab appears to cool relatively linearly 
from 930~K to 665~K over 600 days, while SN~1998S cooled quickly from 1140~K to
900~K over 128 days (between day 227 and day 355), and then more slowly down to 
810~K over the next year.  Apparently the observation of nearly identical 
spectra/temperatures in SN~1998S and SN~1997ab noted by \citet{gerardy00} was 
merely a coincidence as these objects are clearly evolving, albeit over a longer
time period than the previously observed NIR excesses in supernovae.

The total infrared fluxes, calculated by integrating the $H-K$ temperature fits 
over all wavelengths, are also listed
in Table~\ref{singletemp} and plotted in the lower panels of 
Figure~\ref{templum}.  These luminosities should be regarded as lower limits
on the total infrared luminosities, as the near-IR is only sensitive to the 
hottest dust and consequently a significant flux from much colder dust could be 
missed in our data.  

Although SN~1995N showed little temperature evolution, there is significant 
evolution of its infrared luminosity.  For the first four epochs, the luminosity
seems relatively constant at about $7 \times 10^{41}$ erg~s$^{-1}$.  Then after 
day 1000 the luminosity appears to decline exponentially with an e-folding time 
$\approx 600$~d.  While no plateau phase was observed in SN~1997ab, the 
infrared luminosity also appears to decay exponentially, in this
case with an e-folding time of $\approx 450$~d.  However, the NIR behavior of 
SN~1998S was somewhat different.  The inferred IR luminosity of 
SN~1998S increased $\sim 50\%$ between day 260 and day 355, and then 
faded between day 355 and day 796 at a rate similar to that of SN~1997ab.  

While we interpret the IR continuum as thermal emission from dust, it is 
important to estimate the extent that free-free emission might contribute to
the NIR continuum as there are strong recombination lines observed in these 
supernovae. \citet{fransson01} found that in SN 1995N, H$\alpha$ was $\approx 8$
times stronger than expected for standard \ion{H}{1} recombination due to 
collisional excitation and a high optical depth for H$\alpha$.  However, the
higher order Balmer lines were consistent with Case B recombination. Assuming
standard nebular conditions and Case B recombination, the H$\beta$ emission 
per unit volume is $4 \pi j_{H\beta} \approx 1.2 \times 10^{-25} n^2$ erg
cm$^{-3}$ s$^{-1}$ \citep{osterbrock89}.  Using an abundance ratio of He to
H of 0.35 by number \citep{fransson01} the free-free emission
per unit volume $\epsilon_{ff} \approx 3.2 \times 10^{-27} T^{1/2} n^2 $ erg
cm$^{-3}$ sec$^{-1}$ \citep{rybicki79}, where we have assumed a velocity 
averaged Gaunt factor $\bar{g}_B \approx 1.2$.  For $T=10^4$ K, this implies
that $\epsilon_{ff} \approx 3 \epsilon_{H\beta}$.  \citet{fransson01} report
an H$\beta$ flux of $8.8 \times 10^{-15}$ erg s$^{-1}$ cm$^{-2}$ from SN~1995N
on day 718, which translates to a free-free luminosity of $L_{ff} \approx 2 
\times 10^{39}$ erg s$^{-1}$.  Compared to the IR luminosity on day 730 of 
$8 \times 10^{41}$ erg s$^{-1}$, the free-free continuum is down by two 
orders of magnitude and thus does not contribute significantly to the 
observed NIR continuum.

\subsection{Comparison to X-Ray Emission}
Comparison of the observed X-ray and infrared light curves can be used to 
provide clues into
the origin of the IR emission.  Both SN~1995N \citep{fox00} and  SN~1998S 
\citep{pooley01} have been detected in X-rays and the observed X-ray
luminosities are plotted as stars in Figure~\ref{templum}.

One possible dust heating scenario is that the IR emission is from a precursor 
region in the circumstellar medium (CSM) just ahead of the shock 
front that is being heated by X-rays from the post shock gas. In such a 
situation, where the light-travel time is small, we might expect the X-ray 
luminosity and the IR luminosity to be well correlated.  What we find, 
however, is that the IR evolution does not follow the X-ray light curve in
SN~1995N, as the X-ray flux shows no signs of fading nearly 200 days after
the IR begins to fall. Furthermore, the IR flux is significantly 
greater (as much as a factor of five) than the X-rays.  This would require that 
the actual emitted X-ray flux is much larger and that nearly all of the X-rays 
are being absorbed by the circumstellar dust and re-emitted in the IR.  
However, \citet{fransson01} found no indication of strong reddening in the 
optical spectra of SN~1995N, suggesting that the dust extinction in the CSM is 
low.  These problems argue against an IR precursor interpretation.

We note that large IR/X-ray flux ratios are seen in supernova remnants with  
observed ratios varying from $\sim 2$ to $\sim 100$ \citep{dwek87}.  In
remnants the IR emission is from hot, dusty, shocked gas where thermal emission
from collisionally heated dust can be the dominant cooling mechanism. 
However, the IR emission from supernova remnants is typically much cooler
and it seems unlikely that a significant amount of dust could be collisionally 
heated up to 800 -- 1000~K without destroying the dust grains.  Collisional 
heating of dust to $\sim 1000$ K would also require prohibitively large ion 
densities ($ > 10^{21}$ cm$^{-3}$ for $T_e = 10^7$ K; N. Evans 2002, private 
communication).  Furthermore, while there could be some time
lag between changes in the X-ray flux and changes in the IR flux, we would
not expect to see the luminosity drop in the IR before it does in the X-rays as 
seen in SN~1995N.

\subsection{Infrared Echo Models}
On the other hand,
the shape of the SN~1995N IR light curve is similar to the luminosity evolution
of simple infrared echo models.  A number of authors have presented models
describing the infrared echos of supernovae in spherically symmetric dust
distributions \citep{bode80,dwek83b,graham83,graham86}.  In this paper,
we shall adopt the formalism of \citet[EC88 hereafter]{emmering88} which 
adds the generality of non-spherical flattened dust distributions.  A common 
feature of spherical 
models is that they predict an infrared light curve that rises quickly
(only while the supernova is at its brightest) and then levels off in a plateau
before finally decaying.  

The light curve plateau results because the supernova flash clears out a
large dust-free cavity in the center of the circumstellar nebula by vaporizing 
all the dust.  At the edge of this cavity is the hottest and brightest 
dust emission. Outside the cavity, the dust is cooler and
fainter as the supernova flash is spread out over a larger surface.  
For a pulse (delta function) supernova light curve that explodes at time 
$t^\prime$, the observed flux at any time $t$ will come from dust lying on a 
paraboloid described by $t=t^\prime + \frac{r}{c}(1 - \cos \theta)$.  As time 
proceeds, the 
paraboloids become wider and the vertex recedes from the supernova.  At the 
time $t_v=2 r_v/c$, the vertex of the paraboloid will reach the edge of the 
edge of the cavity.  Before this time, the IR luminosity, dominated by emission 
from the cavity's edge, will be nearly constant forming the plateau in the 
IR light curve.  After the paraboloid leaves the cavity, the emission will come
from the fainter and cooler dust and thus the IR luminosity will fade.  A 
plateau is also
expected in models with more realistic SN light curves as long as the SN fades
in significantly less time than it takes for the echo paraboloid to reach
the cavity wall.

The observed IR light curve for SN~1995N is at least qualitatively similar
to the light curves of spherically symmetric infrared echos.  
However, the evolution timescale for SN~1995N is much
longer than for the models that were used to fit infrared echos to previous IR 
excesses in supernovae.  In SN~1979C, SN~1980K and SN~1982E, the IR excesses 
were detected a couple hundred days after maximum light, and the IR light curve
was already declining.  The time $t_v$ at which the re-emission paraboloids 
reached the cavity wall in the model fits was typically $\sim$ 50 -- 100 days.
For SN~1995N this would have to be more like 1000 days where the
observed IR light curve begins to decay rapidly.  This time scale is set by the
radius of the dust-free cavity, which scales as the square root of the supernova
luminosity.  Thus, the much longer time scale would imply either that the
dust around SN~1995N is much easier to vaporize or that SN~1995N is much 
brighter ($\sim 100 \times$) than those other objects.  A more 
likely solution might be that some event prior to core collapse cleared out
the cavity.

\subsubsection{Numerical Calculations}
To test the infrared echo scenario, we have calculated the total IR luminosity 
evolution and near-IR spectra for echo models based on EC88.  These models
assume an axis-symmetric dust distribution where the isodensity curves are 
ellipsoids of revolution where the dust distribution has a $b^{-2}$ radial
dependence.  The circumstellar dust is assumed to have an emission/absorption
efficiency with a power-law wavelength dependence 
$Q_\lambda \propto \lambda^{-n}$.  The model dust cloud has an inner dust-free
zone, as discussed above, and has no sharp outer edge.  The models implicitly 
assume that the optical depth of the dust cloud is low and do not account for 
absorption inside the circumstellar dust nebula, either between the SN and the 
heated dust or 
after re-emission.  Our calculations for the total infrared luminosity 
$F_{\rm IR}(t)$, and the infrared spectrum $F_\lambda(t)$ were performed by 
numerically integrating Equations~8 and 15 of EC88.

The IR dust echo models have a number of parameters that can be varied.  These 
include the
dust emission/absorption efficiency exponent $n$, the size of the inner cavity 
$r_v$, the dust temperature at the inner surface of the cavity $T_v$, the 
SN light curve, the degree to which the dust distribution is flattened, and 
for non-spherical dust distributions the angle between the axis-of-symmetry 
and the line of sight.  Since the observed plateau in the IR light curve was 
suggestive of a spherically symmetric dust distribution, we used spherically 
symmetric echo models for SN~1995N.  This eliminates two of the model parameters
and also makes the numerical calculations easier.  

In principle, if the dust-free cavity were created by the supernova
flash vaporizing the innermost dust, the cavity radius $r_v$, the peak 
luminosity of the supernova, and the dust temperature at the inner edge of the
cavity $T_v$ would all be related. If the dust has a vaporization temperature 
$T_v$, then $r_v$ would be the radius at which the maximum luminosity of 
the supernova would heat the dust to $T_v$.  In practice, however, a number of 
other details about the dust distribution need to be known to relate these
parameters (e.~g., grain size, composition, and absolute values of the 
absorption/emission efficiency and the dust density).  Furthermore, if the 
cavity is the result of pre-explosion evolution and not vaporization due to the 
SN flash, then no relationship would exist.  Therefore, we treat $T_v$ and $r_v$
as independent parameters.  

It is convenient to parameterize the cavity size by
$t_v= 2 r_v/c$, the epoch at which we first observe light re-emitted by dust 
on the cavity surface directly behind the supernova along the line of sight.
This parameter sets the timescale for the infrared evolution. For models
where the supernova light curve decays quickly compared to this timescale, 
$t_v$ is near the ``kink'' in the IR light curve, where the plateau ends and the
dust emission begins to decay.  For SN~1995N, the kink in the IR light curve
occurs near $t=1000$~d (see Fig.~\ref{templum}) so we consider models with 
$t_v$ near this epoch.   

\subsubsection{Short-Flash Echo Model}
\begin{figure*}
\epsscale{.70}
\plotone{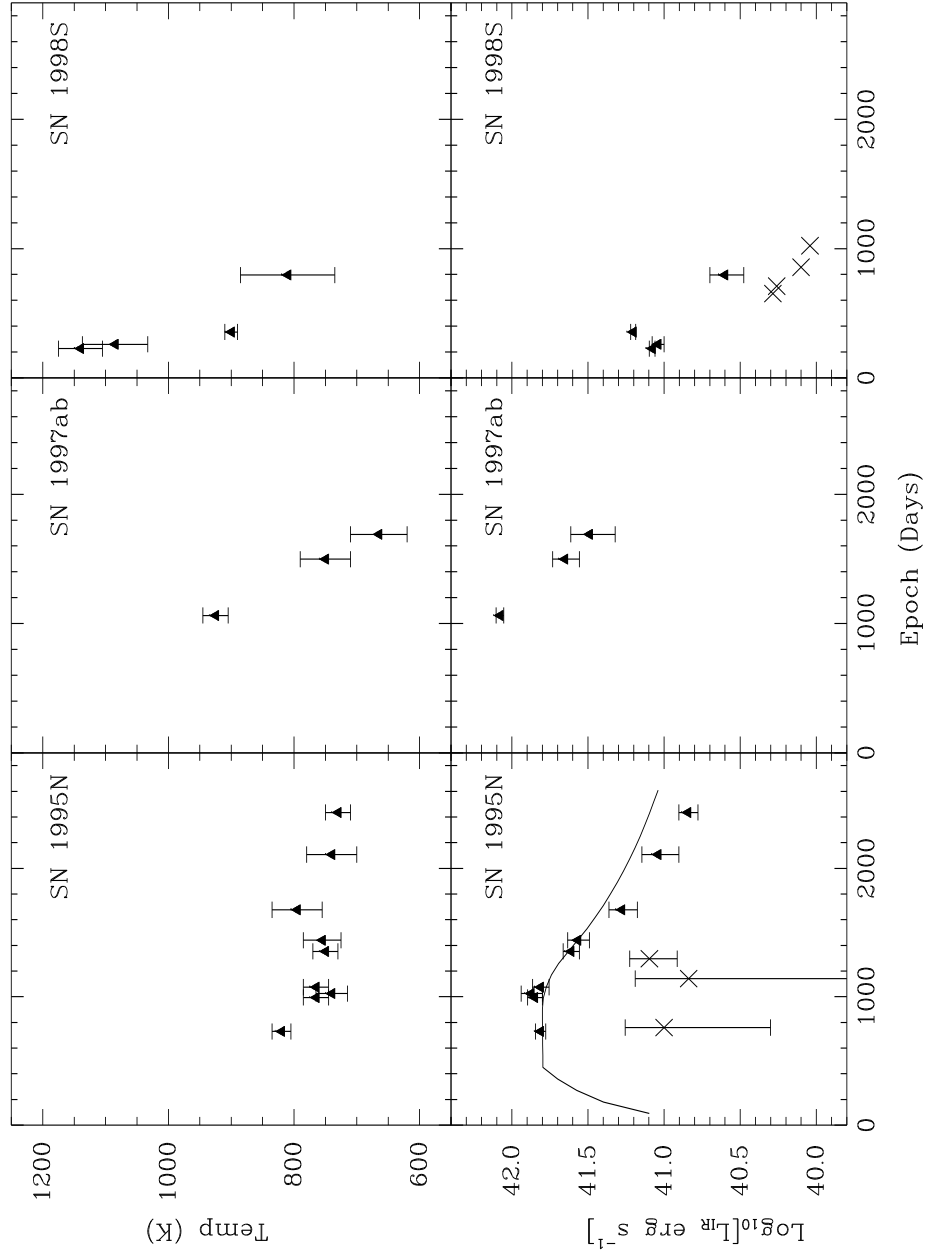}
\caption{The upper panels show the observed $H-K$ temperature evolution for
SN~1995N, SN~1997ab, and SN~1998S, assuming a $\lambda^{-1}$ absorption/emission
efficiency for the dust.  The lower panels plot the inferred IR luminosities 
from the $H-K$ temperature fits (triangles) and observed X-ray luminosities 
(crosses). The solid line in the lower SN~1995N plot is the IR luminosity from
an IR echo model with a short exponential flash as the input supernova light
curve. \label{templum}}
\end{figure*}

Since we have more data on SN~1995N than on the other four Type~IIn SNe, 
we use this 
event to test the IR echo scenario.  The solid line in the lower left panel of 
Figure~\ref{templum}\ shows the 
IR light curve from a model where the input SN light curve is a short flash 
compared to the echo timescale $t_v$.  For the model shown, the SN light curve 
was exponentially decaying with an e-folding time $\Gamma=t_v/10$ and  
dust emission/absorption exponent $n=1$.  
For short-flash models where the supernova light curve fades much faster than 
$t_v$, the results are quite insensitive to the exact details of the input light
curve and a short ``top-hat'' light curve would have produced essentially the 
same results.  

The model IR luminosity was matched to the observed plateau 
luminosity with $t_v$ scaled to 950~d to match the observed kink in
the IR light curve.  As can be seen in Figure~\ref{modres}, the resulting light 
curve is qualitatively similar to the observed IR luminosity evolution of 
SN~1995N but it fades slower after the kink.  However, this is not a strong 
test of the dust echo model. The ``observed'' IR
luminosity is an extrapolation of the single temperature continuum fit to
the $H-K$ color and thus is only sensitive to the hottest dust. In contrast, the
model light curve shows emission from a superposition of dust over a wide
range of temperatures.  As the echo ages and the emission comes from farther
out in the nebula, the IR emission becomes more and more dominated by cooler
dust that is missed by our near-IR photometry.

A better test is to directly compare the observed flux-calibrated spectra to 
the evolution of the model spectrum.  The echo model calculates spectra for each
epoch in dimensionless wavelength units $\lambda$/$\lambda_v$ where 
$\lambda_v \equiv kT_v/h$.  To make a fit for a given set of parameters, the 
model spectrum is 
matched to a single epoch of the data. The model is shifted vertically to match 
the flux levels and horizontally match the shape of the observed spectrum, 
essentially fitting $T_v$.  The same shifts are then applied to the other 
epochs and the resulting spectral evolution can be compared to the data.  

The results of such a procedure are plotted in Figure~\ref{modres}.  In the upper panel we plot the observed spectra of 
SN~1995N with spectra from the short-flash echo model for the corresponding 
epochs.  The model luminosity and $T_v$ have been fit to the second spectrum 
on day 1078.  The cavity time $t_v$ for this model is 950~d and $T_v 
\approx 1100$~K.
The resulting spectral evolution compares well with the observations, although 
the model luminosity fades more quickly.  This is in contrast to the plot
in the lower left panel of Figure~\ref{templum}\ where the model fades more 
slowly 
than the ``observed'' IR luminosity.  Again, the reason for the discrepancy
is that the ``observed'' IR luminosities are only extrapolations based on the
$H-K$ temperature fits and are therefore missing a significant component of 
emission from cool dust.  Since the spectral evolution provides a much more 
direct test, we use this as our primary model discriminant. 

While the short-flash echo model is successful enough to suggest that the IR 
echo scenario may be correct, it involves an unrealistic SN light curve. 
This model describes an echo where most of the SN energy is released quickly, in
this case within the first 100 days.  This timescale, while typical for most 
supernovae, is not really a good description of a Type~IIn supernova like 
SN~1995N.  SN~1995N was observed to have a very slowly decaying light curve 
\citep{baird98}, a feature seen in several other Type~IIn 
events (SN~1987F, SN~1978K, SN~1988Z: \citealt{filippenko97}; SN~1997ab: 
\citealt{hagen97b}).  Also, bright X-ray emission was observed in SN~1995N
at very late times, even beyond the beginning of the observed IR decline.
Both of these facts suggest that a much longer timescale should be used for
the light source of the echo model. 

A longer source light curve is also easier to reconcile with the very large
IR luminosity at late time.  A simple interpolation of the  ``observed'' 
infrared luminosities shown in Figure~\ref{templum}\ yields a total output of 
about $5 \times 10^{49}$ erg in between
days 730 and 2435.  Integration of the model light curve yields a value closer 
to $10^{50}$ erg.  Thus the energy released in the infrared is significantly 
larger than a typical $\sim 100$ day supernova flash, but is perhaps consistent
with the long lasting interaction powered emission seen in other bright Type~IIn
supernovae.  \citet{aretxaga99} found a lower limit of $2 \times 10^{51}$~erg 
for the emission from SN~1988Z, during 8.5 years of observation.  Similar 
emission from SN~1995N would certainly provide enough energy to power the 
observed dust emission, but would again call for a much longer light curve for
the echo model.

\subsubsection{CSM Interaction Echo Model}
While a short-flash model is relatively successful at matching the observed NIR
spectral evolution,
the input SN light curve was not a satisfactory description of SN~1995N and we
therefore examined echo models with longer supernova light curves.  The 
late-time X-ray flux appears to be relatively constant thereby making a long 
``top-hat'' light curve an attractive possibility.  For a long top-hat model, 
the IR 
continuum continues to rise until after the SN light source shuts off.  If the 
cutoff occurs prior to $t_v$ then the IR luminosity levels off until after 
$t_v$, and then begins to fade. Conversely, if the cutoff epoch is after $t_v$, 
then the 
IR flux continues to rise until the light source shuts off and then the IR 
flux quickly decays.

However, what we observe in SN~1995N is not consistent with such a model.
The NIR flux faded very slowly prior to the $t \approx 1000$~d 
kink and fairly rapidly thereafter with no sign of an X-ray cutoff 
$\approx 200$ days after the IR kink.  Furthermore, the qualitative results 
from the long top-hat model hold true for any fairly flat input light-curve.  
Thus the rapid decline of the IR flux after the kink forces us to the 
conclusion that most of the total emission from SN~1995N was released well
before $t=1000$~d even though the late-time X-ray emission was still quite 
bright.  

\begin{figure*}
\epsscale{.70}
\plotone{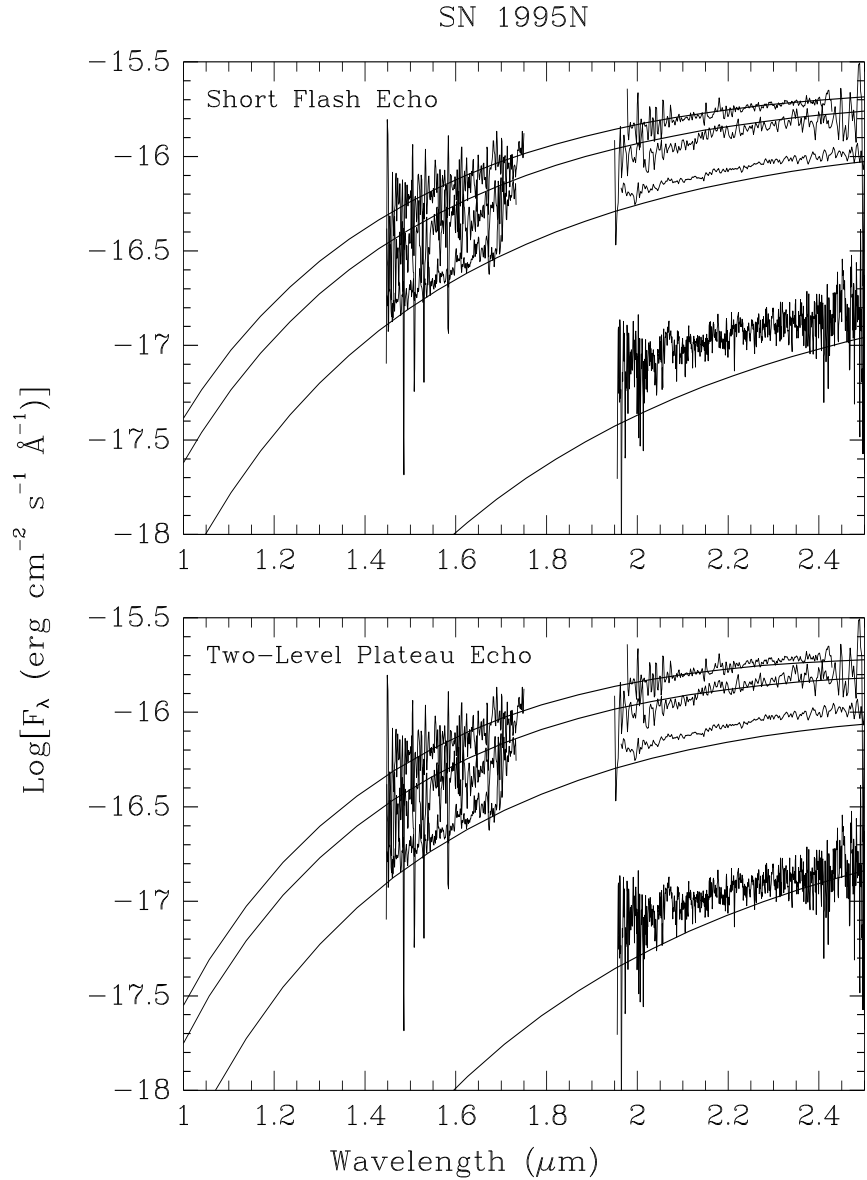}
\caption{Observed near-infrared spectra of SN~1995N compared to calculated 
spectra from infrared echo models.  For the model in the top panel the input
SN light curve was a short exponential decay.  The model in the bottom panel 
results from a two-level ``top-hat'' model with a long bright plateau followed
by a fainter constant source.  The epochs of the observed data are 640, 1078,
1319, and 2435 days, from top to bottom. \label{modres}}
\end{figure*}

This suggests a two-component light curve which we model as a top-hat plus
constant late-time emission.   The results of such a model are plotted in the
bottom half of Figure~\ref{modres}.  For the model shown, the 
absorption/emission exponent $n = 2$, and the echo timescale parameter 
$t_v=800$~d which implies a dust-free cavity radius $r_v \approx 0.3$~pc.  The 
model supernova light curve consists of a bright ``top-hat'' plateau that lasts 
for 400 days, followed by constant luminosity emission 20 times fainter than 
the bright plateau.  The fit value of $T_v \approx 900$~K for this plot.

Figure~\ref{modres}\ shows that this model reproduces the observed IR 
spectral evolution at least as well as the short-flash echo model.  However, 
where the short-flash model was based on an unsatisfying description of the 
SN light curve, the two-component model is based on a more reasonable 
approximation for the UV/X-ray light curve of SN~1995N. What this shows is that
the IR excess observed in SN~1995N can plausibly be explained as an 
infrared echo.  In this case the echo is not powered by the flash of the 
supernova at maximum light but rather by the much more slowly evolving emission
from the circumstellar interaction.  We stress, however, that the particular 
model fit shown in Figure~\ref{modres}\ is probably not unique.  
There are enough parameters that one could likely fit a number of echo 
models to the data.  

Because we did not have nearly as much data from the other Type~IIn supernovae, 
we did not attempt to fit specific echo models for them. But the infrared 
behavior of all of these objects can probably be explained by an interaction 
echo model.  The very large IR luminosities of SN~1997ab and SN~1999Z are 
easiest to explain with such a model.   In SN~1999Z, the rising IR flux seen 
between day 296 and 465 would suggest that the light source was still emitting a
significant amount of energy in this time period.  Since we only observed a 
declining IR flux in SN~1997ab, this would suggest that we caught the IR
emission in SN~1997ab after $t_v$ in the fading era of the echo.  

The behavior of SN~1998S is not suggestive of a spherically symmetric dust 
echo, but is more consistent with an echo from a highly flattened dust 
distribution.  Echo models for flattened dust nebulae with the line of sight 
near the equatorial plane do predict a bump in the IR light curve near $t_v$ 
(EC88).  This would be in line with other evidence for deviations from 
spherical symmetry in SN~1998S.  Spectropolarimetry indicated
that both the SN ejecta and the circumstellar environment were aspherical
\citep{leonard00,wang01}.  Also a well-resolved triple-peak line profile
observed for the hydrogen and helium lines of SN~1998S was interpreted by 
\citet{gerardy00} as emission from
interaction with dense, clumpy circumstellar gas in a ring or disk surrounding
the supernova and probably observed from a fairly high inclination.   

\section{Conclusions}
We have presented near-infrared photometry and spectra from five recent Type~IIn
supernovae that exhibited large late-time infrared excesses.  The infrared
excesses in these objects seem to have a much slower evolution than those 
that have been previously presented in the literature, remaining bright 
many years after maximum light.  The resulting energy release via IR dust
emission is quite large, approximately $0.5$--$1 \times 10^{50}$ erg from 
SN~1995N.  The long IR light curve and large luminosity
cannot be reconciled with emission from newly formed dust in the ejecta.  In 
addition, for
SN~1995N and SN~1998S, the IR luminosity is significantly larger than the 
late-time X-ray emission, ruling out emission from a shock precursor.  
IR emission from collisionally excited dust in shocked gas also seems unlikely 
as the observed temperatures are fairly high ($\sim 1000$~K). Moreover, in 
SN~1995N the
IR flux declines rapidly after day 1000, while the X-ray flux shows little sign 
of fading even $\approx 200$ days later.  

On the other hand, the data do appear to be consistent with an infrared echo of 
the bright, 
slowly-fading emission from the circumstellar interaction.  Simple IR echo
models are able to reproduce the observed near-infrared spectral evolution 
of SN~1995N and such a scenario helps explain the slow evolution and large
IR luminosities seen in these objects.  In principle, detailed studies of 
the IR echos of these objects could provide important information about the 
circumstellar environment of Type~IIn supernovae, and might lend some insight
into their progenitors. In practice, however, to provide significant 
constraints the 
supernova will have to be very well observed, including early time UV and 
X-ray observations, as well as good near- and mid-infrared observations of the
echo.

\acknowledgments
We would like to thank the observatory staffs at MDM, FLWO, KPNO, APO, MMT, 
IRTF,
and Subaru for their excellent support, especially Dr. Hiroshi Terada who was 
enormously helpful during the Subaru observations.  C.~L.~G. and 
R.~A.~F.'s research is supported by NSF grant 98-76703.  K.~N. has been supported 
in part by the Grant-in-Aid for Scientific Research (07CE2002, 12640233) of the
Ministry of Education, Culture, Sports, and Technology in Japan.

\clearpage

\begin{deluxetable}{cccccc}
\tablecaption{Near-Infrared Spectroscopy of SN~1995N\label{nirspec}}
\tablewidth{0in}
\tablehead{
\colhead{Date} & \colhead{JD} & \colhead{Epoch\tablenotemark{a}} &
\colhead{Wavelength} & \colhead{Exposure} &
\colhead{Observatory/Instrument}\\
\colhead{}     & \colhead{}   & \colhead{(days)}                 &
\colhead{Coverage (\micron)}  & \colhead{(sec)}                  & \colhead{}}
\startdata
04 Apr 1996 & 2450179 & 640  & 1.45--1.75 & 1800 & MMT + FSPEC \\*
            &         &      & 1.98--2.43 & 3360 &               \\
05 Apr 1996 & 2450180 & 641  & 1.45--1.75 & 3840 & MMT + FSPEC \\*
            &         &      & 1.98--2.43 & 1920 &             \\
16 Jun 1997 & 2450617 & 1078 & 1.45--1.73 & 3600 & KPNO 4m + CRSP \\*
            &         &      & 1.96--2.49 & 1200 &                \\
12 Feb 1998 & 2450858 & 1319 & 1.44--1.72 & 900  & KPNO 4m + CRSP \\*
            &         &      & 1.92--2.50 & 900  &                \\
05 Mar 2001 & 2451974 & 2435 & 1.93--2.49 & 3600 & Subaru + IRCS  \\
\enddata
\tablenotetext{a}{Adopted date for zero epoch: 04 Jul 1994}
\end{deluxetable}

\clearpage

\begin{deluxetable}{ccccccc}
\tablecaption{Near-Infrared Photometry\label{nirphot}}
\tablewidth{0in}
\tablehead{
\colhead{Date} & \colhead{JD} & \colhead{Epoch\tablenotemark{a}} & \colhead{\textit{J}} & 
\colhead{\textit{H}} & \colhead{\textit{K}} & \colhead{Observatory}}
\startdata
\sidehead{\bf{SN~1995N}}
04 Jul 1996 & 2450269 & 730  & $16.90 \pm 0.03$ & $15.55 \pm 0.03$ & $13.41 \pm 0.04$ & IRTF \\
25 Mar 1997 & 2450533 & 994  & $17.51 \pm 0.11$ & $15.90 \pm 0.05$ & $13.55 \pm 0.05$ & FLWO \\
25 Apr 1997 & 2450564 & 1025 & $17.35 \pm 0.13$ & $16.07 \pm 0.05$ & $13.61 \pm 0.07$ & FLWO \\ 
12 Jun 1997 & 2450612 & 1073 & $17.57 \pm 0.12$ & $15.99 \pm 0.07$ & \nodata          & APO \\
15 Jun 1997 & 2450615 & 1076 & \nodata          & \nodata          & $13.64 \pm 0.05$ & FLWO \\
19 Mar 1998 & 2450892 & 1353 & \nodata          & $16.63 \pm 0.07$ & $14.21 \pm 0.04$ & FLWO \\
10 Apr 1998 & 2450914 & 1375 & $>18.87$ & $17.29 \pm 0.16$ & $14.26 \pm 0.09$ & FLWO \\
14 Jun 1998 & 2450979 & 1440 & \nodata          & $16.70 \pm 0.10$ & $14.30 \pm 0.04$ & FLWO \\
06 Feb 1999 & 2451216 & 1677 & \nodata          & $17.05 \pm 0.11$ & $14.82 \pm 0.07$ & FLWO \\
29 May 1999 & 2451324 & 1785 & \nodata          & $>18.14$ & $14.99 \pm 0.08$ & FLWO \\
11 Apr 2000 & 2451646 & 2107 & $19.57 \pm 0.23$ & \nodata          & $15.69 \pm 0.08$ & MDM \\
12 Apr 2000 & 2451647 & 2108 & \nodata          & $18.15 \pm 0.12$ & \nodata          & MDM \\
14 Apr 2000 & 2451649 & 2110 & \nodata          & \nodata          & $15.85 \pm 0.10$ & FLWO \\
05 Mar 2001 & 2451974 & 2435 & \nodata		& $18.82 \pm 0.06$ & $16.31 \pm 0.06$ & Subaru \\
30 Mar 2001 & 2452000 & 2461 & \nodata          & \nodata          & $16.41 \pm 0.05$ & MDM \\
01 May 2001 & 2452032 & 2493 & \nodata		& \nodata	   & $16.47 \pm 0.14$ & MDM \\ 
\sidehead{\bf{SN~1997ab}}
09 Mar 1999 & 2451248 & 1062 & $17.58 \pm 0.05$ & $15.67 \pm 0.03$ & $13.86 \pm 0.04$ & MDM \\
18 May 2000 & 2451684 & 1498 & \nodata          & $17.96 \pm 0.13$ & $15.54 \pm 0.04$ & MDM \\
26 Nov 2000 & 2451876 & 1690 & \nodata          & $19.30 \pm 0.18$ & $16.48 \pm 0.06$ & MDM \\
\sidehead{\bf{SN~1998S}}
20 Jun 1998 & 2450986 & 94   & $14.37 \pm 0.03$ & $14.07 \pm 0.05$ & $13.46 \pm 0.07$ & MDM \\
31 Oct 1998 & 2451119 & 227  & $16.29 \pm 0.05$ & $14.83 \pm 0.04$ & $13.50 \pm 0.04$ & MDM \\
03 Dec 1998 & 2451152 & 260  & $16.74 \pm 0.08$ & $15.05 \pm 0.06$ & $13.62 \pm 0.07$ & MDM \\
08 Mar 1999 & 2451247 & 355  & $17.26 \pm 0.07$ & $15.46 \pm 0.04$ & $13.57 \pm 0.02$ & MDM \\
09 Mar 1999 & 2451248 & 356  & $17.28 \pm 0.08$ & $15.46 \pm 0.03$ & $13.59 \pm 0.03$ & MDM \\
22 May 2000 & 2451688 & 796  & \nodata		& $17.57 \pm 0.19$ & $15.38 \pm 0.09$ & MDM \\
\sidehead{\bf{SN~1999Z}}
01 Dec 1999 & 2451515 & 296  & $17.71 \pm 0.05$ & \nodata          & $16.49 \pm 0.05$ & MDM \\
18 May 2000 & 2451684 & 465  & $18.25 \pm 0.08$ & \nodata          & $16.07 \pm 0.07$ & MDM \\
23 Nov 2000 & 2451873 & 654  & $>20.11$ & $19.55 \pm 0.23$ & $16.84 \pm 0.06$ & MDM \\
\sidehead{\bf{SN~1999el}}
18 May 2000 & 2451684 & 198  & $18.33 \pm 0.15$ & \nodata          & $16.29 \pm 0.08$ & MDM \\
23 Nov 2000 & 2451873 & 387  & $>18.99$ & $>18.53$ & $16.84 \pm 0.12$ & MDM \\
\enddata
\tablenotetext{a}{Adopted dates for zero epoch --- SN~1995N: 04 Jul 1994; SN~1997ab: 11 Apr 1996; 
SN~1998S: 18 Mar 1998; SN~1999Z: 08 Feb 1999; SN~1999el: 02 Nov 1999.  See text for details.}
\end{deluxetable}

\clearpage

\begin{deluxetable}{lcccccccc}
\tablewidth{0in}
\tabletypesize{\scriptsize}
\tablecolumns{9}
\tablecaption{Single Temperature Fits\label{singletemp}}
\tablehead{
\colhead{Epoch}	& \multicolumn{2}{c}{$Q_{\lambda} \propto 1$} & & 
\multicolumn{2}{c}{$Q_{\lambda} \propto \lambda^{-1}$} & &
\multicolumn{2}{c}{$Q_{\lambda} \propto \lambda^{-2}$} \\
\cline{2-3} \cline{5-6} \cline{8-9}
\colhead{(days)} & \colhead{T$_{d}$ (K)} & \colhead{L$_{d}$ 
($10^{40}$erg~s$^{-1}$)\tablenotemark{a}}  & &
\colhead{T$_{d}$ (K)} & \colhead{L$_{d}$ 
($10^{40}$~erg~s$^{-1}$)\tablenotemark{a}} & &
\colhead{T$_{d}$ (K)} & \colhead{L$_{d}$ 
($10^{40}$~erg~s$^{-1}$)\tablenotemark{a}}}
\startdata
\sidehead{\bf{SN~1995N}}
730	& $930 \pm 20$ & $89 \pm 8$ & & $830 \pm 15$ & $65 \pm 5$ & & 
$745 \pm 15$ & $52 \pm 4$\\
994	& $860 \pm 25$  & $101 \pm 12$ & & $770 \pm 20$ & $71 \pm 8$ & & 
$700 \pm 15$ & $55 \pm 5$\\	
1025	& $830 \pm 30$ & $108 \pm 19$ & & $745 \pm 25$ & $75 \pm 11$ & & 
$680 \pm 20$ & $57 \pm 8$\\
1073,1076 & $860 \pm 30$ & $92 \pm 13$ & & $775 \pm 20$ & $65 \pm 8$ & & 
$700 \pm 20$ & $50 \pm 6$\\
1353	& $840 \pm 25$ & $59 \pm 8$ & & $755 \pm 20$ & $41 \pm 5$ & & 
$685 \pm 20 $ & $32 \pm 4$ \\
1440	& $845 \pm 40$ & $53 \pm 10$ & & $760 \pm 30$ & $37 \pm 6$ &
& $690 \pm 25$ & $29 \pm 4$ \\
1677	& $900 \pm 50$ & $27 \pm 6$ & & $805 \pm 40$ & $19 \pm 4$ &
& $725 \pm 35$ & $15 \pm 3$ \\
2107,2108 & $830 \pm 45 $ & $16 \pm 4$ & & $745 \pm 40$ & $11 \pm 3$ &
& $680 \pm 30$ & $8 \pm 2$\\
2435	& $815 \pm 25$ &$10 \pm 2$ & & $735 \pm 20$ &$7 \pm 1$ &
& $670 \pm 15$ & $5 \pm 0.6$ \\
\sidehead{\bf{SN~1997ab}}
1062	& $1060 \pm 20$ & $156 \pm 11$ & & $930 \pm 15$ &$120 \pm 7$ & & 
$825 \pm 15$ & $100 \pm 5$\\
1498	& $835 \pm 50$ & $65 \pm 14$ & & $750 \pm 40$ &$45 \pm 9$ & & 
$685 \pm 30$ & $35 \pm 6$\\
1690	& $730 \pm 55$ &$48 \pm 17$ & & $665 \pm 45$ & $31 \pm 10$ & & 
$610 \pm 40$ & $22 \pm 6$\\
\sidehead{\bf{SN~1998S}}
227	& $1350 \pm 45$ & $15 \pm 0.8$ & & $1140 \pm 35$ & $12 \pm 0.5$ & & 
$990 \pm 25$ & $ 11 \pm 0.4$\\
260	& $1275 \pm 70$ & $14 \pm 1.5$ & & $1085 \pm 52$ & $11 \pm 1$ & & 
$950 \pm 40$ & $ 10 \pm 0.8$\\
355	& $1025 \pm 15$ & $21 \pm 0.9$ & & $905 \pm 10$ & $16 \pm 0.6$ & & 
$805 \pm 10$ & $ 13 \pm 0.4$\\
796	& $910 \pm 100$ & $5.5 \pm 1.9$ & & $810 \pm 75$ & $4.0 \pm 1$ & & 
$730 \pm 60$ & $3.1 \pm 0.8$\\	
\sidehead{\bf{SN~1999Z}}
654	& $760^{+80}_{-55}$ & $515^{+235}_{-195}$ & & $690^{+65}_{-45}$ & 
$340^{+135}_{-115}$ & & $630^{+55}_{-35}$ & $250^{+90}_{-80}$\\
\sidehead{\bf{SN~1999el}}
387	& $\lesssim 1245$ & $\gtrsim 2$ & & $\lesssim 1060$ &
$\gtrsim 1$ & & $\lesssim 890$ & $\gtrsim 1$
\enddata
\tablenotetext{a}{Quoted sigmas do not include distance uncertainty.}
\end{deluxetable}

\clearpage

\end{document}